\documentclass[modern]{aastex631}

\begin{document}

\title{Updated forecast for TRAPPIST-1 times of transit for all seven exoplanets incorporating JWST data}

\author[0000-0002-0802-9145]{Eric Agol}
\affiliation{Department of Astronomy,
University of Washington,
Seattle, WA 98195, USA}
\correspondingauthor{Eric Agol}
\email{agol@uw.edu}

\author[0000-0002-0832-710X]{Natalie H. Allen}\altaffiliation{NSF Graduate Research Fellow}
\affiliation{Department of Physics and Astronomy, Johns Hopkins University, 3400 N. Charles Street, Baltimore, MD 21218, USA}

\author[0000-0001-5578-1498]{Björn Benneke}
\affiliation{Trottier Institute for Research on Exoplanets and Department of Physics, Universit\'{e} de Montr\'{e}al, Montreal, QC, Canada}

\author[0000-0001-6108-4808]{Laetitia Delrez}
\affil{{Astrobiology Research Unit}, {Universit\'e de Li\`ege}, {{All\'ee du 6 ao\^ut 19}, {Li\`ege}, {4000}, {Belgium}}} 

\author[0000-0001-5485-4675]{Ren\'{e} Doyon}
\affiliation{Trottier Institute for Research on Exoplanets and Department of Physics, Universit\'{e} de Montr\'{e}al, Montreal, QC, Canada}

\author[0000-0002-7008-6888]{Elsa Ducrot}
\affil{LESIA, Observatoire de Paris, CNRS, Universit\'e Paris Diderot, Universit\'e Pierre et Marie Curie, 5 place Jules Janssen, 92190 Meudon, France}
\affil{AIM, CEA, CNRS, Universit\'e Paris-Saclay, Universit\'e de Paris, F-91191 Gif-sur-Yvette, France}

\author[0000-0001-9513-1449]{N\'estor Espinoza}
\affiliation{Space Telescope Science Institute, 3700 San Martin Drive, Baltimore, MD 21218, USA}
\affiliation{Department of Physics and Astronomy, Johns Hopkins University, 3400 N. Charles Street, Baltimore, MD 21218, USA}

\author[0000-0003-0854-3002]{Am\'{e}lie Gressier}
\affiliation{Space Telescope Science Institute, 3700 San Martin Drive, Baltimore, MD 21218, USA}

\author[0000-0002-6780-4252]{David Lafrenière}
\affiliation{Trottier Institute for Research on Exoplanets and Department of Physics, Universit\'{e} de Montr\'{e}al, Montreal, QC, Canada}

\author[0000-0003-4676-0622]{Olivia Lim}
\affiliation{Trottier Institute for Research on Exoplanets and Department of Physics, Universit\'{e} de Montr\'{e}al, Montreal, QC, Canada}

\author[0000-0002-0746-1980]{Jacob Lustig-Yaeger}
\affiliation{JHU Applied Physics Laboratory, 11100 Johns Hopkins Rd, Laurel, MD 20723, USA}

\author[0000-0002-2875-917X]{Caroline Piaulet-Ghorayeb}
\affiliation{Trottier Institute for Research on Exoplanets and Department of Physics, Universit\'{e} de Montr\'{e}al, Montreal, QC, Canada}

\author[0000-0002-3328-1203]{Michael Radica}
\affiliation{Trottier Institute for Research on Exoplanets and Department of Physics, Universit\'{e} de Montr\'{e}al, Montreal, QC, Canada}

\author[0000-0003-4408-0463]{Zafar Rustamkulov}
\affiliation{Department of Earth \& Planetary Sciences, Johns Hopkins University, Baltimore, MD, USA}

\author[0000-0001-7393-2368]{Kristin S. Sotzen}
\affiliation{JHU Applied Physics Laboratory, 11100 Johns Hopkins Rd, Laurel, MD 20723, USA}




\begin{abstract}
The TRAPPIST-1 system has been extensively observed with JWST in the near-infrared with the goal of measuring atmospheric transit transmission spectra of these temperate, Earth-sized exoplanets.   A byproduct of these observations has been
much more precise times of transit compared with prior available data from Spitzer,
HST, or ground-based telescopes.  In this note we use 23 new timing measurements
of all seven planets in the near-infrared from five JWST observing programs 
to better forecast and constrain the future times
of transit in this system.  In particular, we note that the transit times
of TRAPPIST-1h have drifted significantly from a prior published analysis by up to tens of minutes.  Our newer forecast has a higher precision, with median statistical uncertainties ranging from 7-105 seconds during JWST Cycles 4 and 5.
Our expectation is that this forecast will help to improve planning of future
observations of the TRAPPIST-1 planets, whereas we postpone a full dynamical
analysis to future work. 
\end{abstract}



\section{Introduction} \label{sec:intro}

During Cycles 1-3 of JWST, transits of all seven planets in the TRAPPIST-1 system \citep{Gillon2017} have been observed in
the near-infrared with NIRISS-SOSS \citep{Albert2023} and NIRSPEC-BOTS \citep{Jakobsen2022}, as well as four additional transits observed with MIRI F1500W during a phase-curve observation of the system \citep{Gillon2023}.  In this
note we focus on touching-up the timing forecast for this system; hence,
we only utilize near-IR transits thanks to their higher precision. 
From July 2022 - December 2023, using JWST NIRISS
or NIRSPEC, there have been three transits observed of planet b (NIRISS-SOSS, JWST GO-2589, PI: \citealt{Lim2023}; JWST GO-1981, PI: Stevenson/Lustig-Yaeger), 
five of planet c (NIRISS-SOSS, JWST GO-2589, PI: Lim; NIRspec-BOTS, JWST GO-2420, PI: Rathcke), two of planet d (NIRSpec-BOTS, JWST GO-1201, PI: Lafreni\`ere),
four of planet e (NIRSpec-BOTS, JWST GTO-1331, PI: Lewis), five of planet f (NIRISS-SOSS, JWST GO-1201, PI: Lafreni\`ere),
two of planet g (NIRSpec-BOTS, JWST GO-2589, PI: Lim), and two of planet h (NIRSpec-BOTS, JWST GO-1981, PI: Steveson/Lustig-Yaeger).
 
For the unpublished times of transit, the light curves were analyzed using standard pipelines with quadratic limb-darkened transit models \citep{Mandel2002}. 
The first two transit times of planet b were published in \citet{Lim2023}, while the third was simultaneous with a transit of planet h, described below.  For planet c, NIRISS/SOSS light curves were produced using the \texttt{exoTEDRF} code \citep{Feinstein2023, Radica2023, Radica2024exotedrf}, and fitted using \texttt{juliet} \citep{Espinoza2019}, following the same procedure as \citet{Radica2024} for each of the two visits. Planet c NIRSpec PRISM light curves were generated using the \texttt{transitspectroscopy} pipeline \citep{Espinoza_2022} and fitted with the \texttt{juliet} python package \citep{Espinoza2019} to provide best-fit transit timings and their uncertainties.
For planets d, f and g, the light curves were fitted with the same framework as in \citet{Lim2023}. For planet h (and one simultaneous transit of planet b) the transit timings were derived using 
the Eureka pipeline, which was run to produce the light curves, and then the \texttt{trafit} \citep{Gillon2010, Gillon2012} code was used to fit the transits and provide the best fit timings and their uncertainties.  When timing uncertainties are two-sided, we used the greater of the uncertainties to provide a symmetric
error bar for utilizing a chi-square statistic.  When multiple timing analyses were carried out, we checked that the measured times were consistent across analyses, and we used the larger of the timing uncertainties for each transit to make our forecast conservative.

The measured times are for planet b: $9779.210475\pm0.000025$, $9780.72134581\pm 0.000025$, $10289.8849446\pm 0.0000098$;
planet c: $9772.420388\pm 0.000012$, $9881.401521\pm 0.000032$,
$10247.0939505\pm 0.0000125$, $10249.515096\pm 0.000034$, 
$10256.7810661\pm 0.0000185$; planet d: $9889.264477\pm 0.000024$,
$9893.313946\pm 0.000060$;
planet e: $10118.459836\pm 0.000032$, $10124.558969\pm 0.000022$,     
$10148.956261\pm 0.000043$, $10246.539286\pm 0.000030$;
planet f: $9881.035336\pm 0.000064$,
$10111.185753\pm 0.000043$,
$10120.387970\pm 0.000045$,
$10129.593868\pm 0.000096$,
$10148.005153\pm 0.000042$;
planet g:
$9777.8353589\pm 0.0000105$,
$9926.053178\pm 0.000017$; and
planet h: $10139.74928085\pm 0.0000809$ and
$10289.88762392\pm 0.000299$.  Each time is given as $BJD_{TDB}-2,450,000$ in days.

\section{Timing analysis} 

We carried out a transit-timing analysis using the timing data published in
\citet{Agol2021a} as well as the 23 new JWST times listed above. 
The \texttt{NbodyGradient.jl} \citep{Agol2021b} code was used to compute the transit
times and their derivatives with respect to the initial conditions,
using the same integrator, initial conditions, time step, and parameter set described in \citet{Agol2021a}.  The computation
is Newtonian and plane-parallel.  We
minimized the chi-square of the fit using the \texttt{LsqFit.jl} 
package which implements a Levenberg-Marquardt optimization algorithm.

With the optimization completed, we integrated the model to July 2027
to cover the end of JWST Cycle 5.  Using the Laplace approximation, we
propagated the uncertainties using the covariance matrix of the
optimum model parameters dotted with the Jacobian of the transit times
with respect to the initial model parameters, yielding the uncertainties on the forecast times.  Yet, these forecast uncertainties almost certainly underestimate the true uncertainty.

Notably, some of the JWST transits were affected by flares, and they may also be impacted by instrumental and/or stellar variability noise.  Moreover, using the Laplace approximation may not fully represent the probability distribution.  To address these issues we did the following: 1).\ we successively dropped a single transit of the 23 new JWST times, and re-optimized the timing model using the remaining data; 2).\ we used each optimized model to forecast the dropped time and its uncertainty; 3).\ we computed the normalized residuals of these 23 JWST times with respect to the forecast time and uncertainty based on the remaining data.  We found that the cumulative distribution of the normalized residuals is consistent with a Gaussian distribution, but with the uncertainties inflated by a factor of 3.14.  Consequently, for our forecast times based on the entire dataset, we inflate the forecast uncertainties by a factor of 3.14; this approach is analogous to leave-one-out cross-validation.
We leave a more detailed
analysis, including a markov-chain monte carlo analysis with a more
robust likelihood function, to future work.

\begin{figure}[ht!]
\plotone{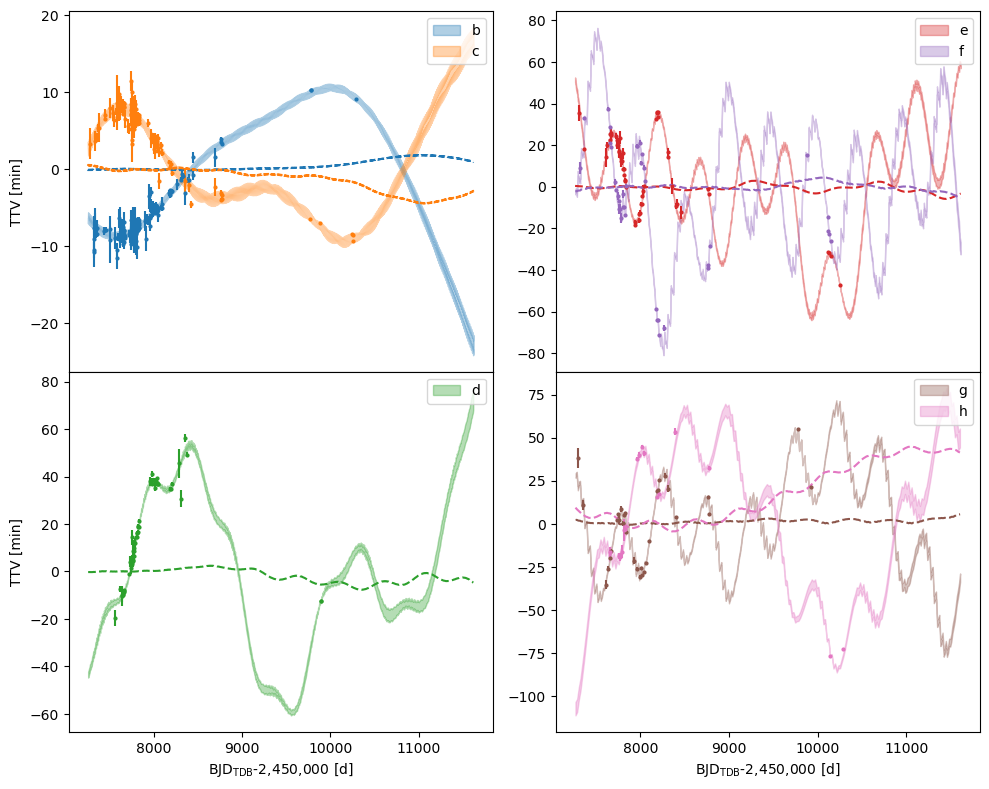}

\caption{Transit-timing variations of the seven TRAPPIST-1 planets.  Error bars are measurements from Agol et al. (2021), and the JWST times from this paper.  Dashed curves show the difference between the new timing solution minus that from Agol et al. (2021).  The data behind the figure are available for quantitative forecasting purposes.
\label{fig:T1_JWST}}
\end{figure}

Figure \ref{fig:T1_JWST} shows the resulting transit-timing variations
as a function of time.  
The deviation of the new timing solution relative to the best-fit timing solution
presented in \citet{Agol2021a} is shown with dashed lines in each panel.
In all cases the agreement is excellent through the end of 2019, while
the timing solutions diverge slightly over the next few years.  The
most extreme case of divergence is for planet h, which is already
arriving $\ga$0.5 hr later than the forecast from \citet{Agol2021a}.  The
other planets have forecasts accurate to $\approx$minutes through July 2027.

The values of the forecast times can be found in the data behind the figure.
A Jupyter notebook used to produce this code is available from the first author.  As a check on the forecast times, we compared with five unpublished transit times from July 2024 for planets b, d and e under JWST program GO-6456 (PI: Natalie Allen and Néstor Espinoza), and all five lie within 1-$\sigma$ of our forecast.

Several questions arise: Can we better constrain the masses and
orbital parameters when including the JWST data?  Is there evidence
for an 8th planet, a moon, or other non-Keplerian effects?  How
much better can we constrain the bulk densities of these planets using JWST data?
To address these questions in detail may require more data for redundancy,
as well as a careful analysis of the JWST data in light of the
impact of frequent flares found in TRAPPIST-1 with JWST \citep{Howard2023}
and the possible presence of correlated noise.  Hundreds
of unpublished transit measurements from ground-based telescopes (Ducrot, private
communication), as well as observations of the secondary eclipses
of planets b and c with MIRI \citep{Greene2023,Zieba2023}, could be included to further improve a dynamical analysis.  For now, though, our forecast transit
times may be used for future planning purposes to attempt spectroscopic
measurements with JWST in the presence of stellar inhomogeneities \citep{Lim2023,deWit2024},
while in turn these measurements will help to further constrain the
dynamics of this system.  For planning purposes for secondary
eclipses, it should suffice to take the half-way point between
adjacent transits (this ignores the light travel time and
eccentricity offset, but both should be insignificant).

\begin{acknowledgments}
E.A. acknowledges support from NSF grant AST-1907342, NASA NExSS grant No.\ 80NSSC18K0829, and NASA XRP grant 80NSSC21K1111.
O.L. acknowledges support from the Fonds de recherche du Qu\'{e}bec --- Nature et technologies (FRQNT). N.H.A. acknowledges support by the National Science Foundation Graduate Research Fellowship under Grant No. DGE1746891. M.R.\ acknowledges financial support from the Natural Sciences and Engineering Research Council of Canada (NSERC) and FRQNT. C.P.-G. acknowledges support from the NSERC Vanier scholarship, and the Trottier Family Foundation. This work was funded by the Institut Trottier de Recherche sur les Exoplan\`{e}tes (iREx).
E.D. acknowledge support from the Paris Observatory-PSL fellowship. 
\end{acknowledgments}

%

\vspace{5mm}
\facilities{JWST (NIRISS, NIRSpec).}


\software{\texttt{NbodyGradient.jl} \citep{Agol2021b},  
          \texttt{juliet} \citep{Espinoza2019}, \texttt{exoTEDRF} \citep{Feinstein2023,Radica2023,Radica2024}, \texttt{transitspectroscopy} \citep{Espinoza_2022}, \texttt{trafit} \citep{Gillon2010,Gillon2012}, \texttt{LsqFit.jl} (\url{https://github.com/JuliaNLSolvers/LsqFit.jl}}




\bibliography{Agol_T1_JWST_forecast}{}
\bibliographystyle{aasjournal}



\end{document}